\documentclass[graybox, vecphys]{svmult}

\usepackage{type1cm}        
\usepackage{makeidx}         
\usepackage{graphicx}        
\usepackage{multicol}        
\usepackage[bottom]{footmisc}

\usepackage{newtxtext}       %
\usepackage{newtxmath}       

\usepackage{amsmath,amsfonts,thmtools}
\usepackage{dsfont,bbm,epsfig,amstext}
\usepackage[nolist]{acronym}
\usepackage{tikz,pgfplots}
\usepgfplotslibrary{fillbetween}
\usetikzlibrary{shapes,arrows}

\renewcommand{\Im}{\operatorname{Im}}

\newcommand{\setX}{\mathbbmss{X}}

\newcommand{\setR}{\mathbbmss{R}}
\newcommand{\setV}{\mathbbmss{V}}

\DeclareMathOperator*{\argmin}{argmin}

\newcommand{\rmc}{\mathrm{c}}

\newcommand{\rmp}{\mathrm{p}}
\newcommand{\rmq}{\mathrm{q}}

\newcommand{\rmR}{\mathrm{R}}

\newcommand{\rmG}{\mathrm{G}}

\newcommand{\rmv}{\mathrm{v}}

\newcommand{\maH}{\mathcal{H}}
\newcommand{\maF}{\mathcal{F}}

\newcommand{\mae}{\mathcal{E}}
\newcommand{\maz}{\mathcal{Z}}

\newcommand{\mam}{\mathcal{M}}

\newcommand{\mg}{\mathcal{G}}

\newcommand{\bxx}{\mathbf{x}}

\newcommand{\bvv}{\mathbf{v}}

\newcommand{\bgg}{\mathbf{g}}

\newcommand{\bx}{{\boldsymbol{x}}}
\newcommand{\hx}{{\hat{x}}}

\newcommand{\hxx}{\hat{\mathrm{x}}}
\newcommand{\xx}{\mathrm{x}}
\newcommand{\yy}{\mathrm{y}}
\newcommand{\zz}{\mathrm{z}}

\newcommand{\bhx}{{\boldsymbol{\hat{x}}}}

\newcommand{\bz}{{\boldsymbol{z}}}

\newcommand{\bv}{{\boldsymbol{v}}}

\newcommand{\dif}{\mathrm{d}}

\newcommand{\by}{{\boldsymbol{y}}}

\newcommand{\trp}{\mathsf{T}}

\newcommand{\mA}{\mathbf{A}}

\newcommand{\mOmega}{\mathbf{\Omega}}
\newcommand{\mR}{\mathbf{R}}

\newcommand{\mI}{\mathbf{I}}
\newcommand{\mone}{\mathbf{1}}
\newcommand{\mJ}{\mathbf{J}}

\newcommand{\mQ}{\mathbf{Q}}
\newcommand{\mS}{\mathbf{S}}
\newcommand{\mU}{\mathbf{U}}

\newcommand{\mD}{\mathbf{D}}
\newcommand{\mM}{\mathbf{M}}

\newcommand{\mT}{\mathbf{T}}

\newcommand{\Ex}[2]{\mathbbmss{E}_{#2}\left\{#1\right\}}
\newcommand{\norm}[1]{\lVert #1 \rVert}
\newcommand{\set}[1]{\left\lbrace #1 \right\rbrace}
\newcommand{\brc}[1]{\left( #1 \right)}
\newcommand{\dbc}[1]{\left[ #1 \right]}

\newcommand{\abs}[1]{\lvert #1 \rvert}
\newcommand{\tr}[1]{\mathrm{tr} \{ #1 \}}

\makeindex             
\usepackage{bibentry}
\nobibliography* 
\usepackage[resetlabels]{multibib}
\newcites{LG}{References}     
\begin{document}
	\begin{acronym}
	\acro{dsn}[DSN]{distributed sensing network}
	\acro{mmv}[MMV]{multiple measurement vectors}
	\acro{dcs}[DCS]{distributed compressive sensing}
	\acro{mse}[MSE]{mean squared error}
	\acro{awgn}[AWGN]{additive white Gaussian noise}
	\acro{iid}[i.i.d.]{independent and identically distributed}
	\acro{np}[NP]{nondeterministic polynomial time}
	\acro{rls}[RLS]{regularized least squares}
	\acro{rs}[RS]{replica symmetry}
	\acro{rsb}[RSB]{replica symmetry breaking}
	\acro{lasso}[LASSO]{least absolute shrinkage and selection operator}
	\acro{snr}[SNR]{signal-to-noise ratio}
	\acro{map}[MAP]{maximum-a-posteriori}
	\acro{pmf}[PMF]{probability mass function}
	\acro{pdf}[PDF]{probability density function}
	\acro{cdf}[CDF]{cumulative distribution function}
\end{acronym}
	
\title*{Analysis of Sparse Recovery Algorithms via the Replica Method}
\subtitle{
	\normalfont
	A Comprehensive Introduction to the Applications of the Replica Method in Analysis of Large Inference Problems\vspace*{3mm} \\
	\small
	This is the initial version of the contribution to the book ``Compressed Sensing in Information Processing''. The peer-reviewed version has been published in \cite{Bereyhi2022}.
}
\author{Ali Bereyhi, Ralf R. M\"uller and Hermann Schulz-Baldes}
\authorrunning{Bereyhi et al.} 
\institute{Ali Bereyhi \at Institute for Digital Communications, Friedrich-Alexander Universit\"at Erlangen-N\"urnberg, Cauerstrasse 7, 91058, Erlangen \email{ali.bereyhi@fau.de}
	\and Ralf R. M\"uller \at Institute for Digital Communications, Friedrich-Alexander Universit\"at Erlangen-N\"urnberg, Cauerstrasse 7, 91058, Erlangen \email{ralf.r.mueller@fau.de}
\and Hermann Schulz-Baldes \at Department of Mathematics, Friedrich-Alexander Universit\"at Erlangen-N\"urnberg, Cauerstrasse 11, 91058, Erlangen \email{schuba@mi.uni-erlangen.de}
}
%
%
\maketitle

\abstract*{
	This manuscript goes through the fundamental connection between statistical mechanics and estimation theory by focusing on the particular problem of \textit{compressive sensing}. We first show that the asymptotic analysis of a sparse recovery algorithm is mathematically equivalent to the problem of calculating the free energy of a \textit{spin glass} in the thermodynamic limit. We then use the \textit{replica method} from statistical mechanics to evaluate the performance in the asymptotic regime. The asymptotic results have several applications in communications and signal processing. We briefly go through two instances of these applications: Characterization of joint sparse recovery algorithms used in distributed compressive sensing, and tuning of receivers employed for detection of spatially modulated signals.
}

\abstract{
	This manuscript goes through the fundamental connections between statistical mechanics and estimation theory by focusing on the particular problem of \textit{compressive sensing}. We first show that the asymptotic analysis of a sparse recovery algorithm is mathematically equivalent to the problem of calculating the free energy of a \textit{spin glass} in the thermodynamic limit. We then use the \textit{replica method} from statistical mechanics to evaluate the performance in the asymptotic regime. The asymptotic results have several applications in communications and signal processing. We briefly go through two instances of these applications: Characterization of joint sparse recovery algorithms used in distributed compressive sensing, and tuning of receivers employed for detection of spatially modulated signals.
}

\section{Introduction}
Statistical mechanics deals with the analysis of very large many-particle systems and seeks the following ultimate goal: Starting from the \textit{microscopic} behavior of individual particles, it tries to find out the \textit{macroscopic} properties of the system. The system size is, however, so large that it is not possible to solve the microscopic equations of motion. Statistical mechanics follows an alternative approach: It describes the microscopic behavior of the system particles via a stochastic model and extracts the desired \textit{deterministic} properties via statistical analysis.

The goal and techniques of statistical mechanics are in various aspects similar to those of information theory. This connection has been widely investigated in the literature \cite{merhav2010statistical}. In addition to all interesting theoretical aspects of this connection, the links between the two theories lead to a key achievement: The analytical tools of statistical mechanics can be used to address asymptotic analysis in information theory and its applications.

In this manuscript, we use one particular statistical mechanical tool, namely the replica method, to investigate the asymptotic performance of a large class of sparse recovery algorithms. The interest in characterizing the asymptotic performance has several origins: The most natural one is to have an analytic bound on the performance of a given recovery scheme. This is however not the only application. Sparse recovery is used in several other applications in which the asymptotic performance characterization is useful for system design. In Section~\ref{sec:app}, we give two particular instances; namely, the example of deriving error bounds in distributed compressive sensing, and tuning algorithms used for detection of spatially modulated signals.

The focus of this manuscript is on the asymptotic analysis of a generic compressive sensing setting via the replica method. As a result, the contents give a comprehensive overview on the replica method and its applications to asymptotic analyses in communications and signal processing. The discussions in this manuscript can be followed in greater detail in \cite{bereyhi2020thesis}.

\section{A Multi-Terminal Setting for Compressive Sensing}
\label{sec:SysMod}
We consider a generic multi-terminal setting. This setting includes the classical \textit{single-terminal} compressive sensing setting, as well as other scenarios of sparse  recovery.

Consider a \ac{dsn} with $J$ \textit{correlated sparse} source signals, namely $x_j\brc{t}$ for $j\in \dbc{J}$. Here, the notation $\dbc{J}$ is defined as 
\begin{equation}
	\dbc{J} \coloneqq \set{1, \ldots , J},
\end{equation}
and is used through the manuscript to shorten the presentation. The source signals are sampled at the time instances, $t=t_n$ for $n\in \dbc{N}$. Let $\bx_j \in \setX^{N\times 1}$ for $\setX\subset \setR$ denote the vector of samples collected from the $j$-th source signal. We assume that the sampling is performed, such that the \textit{temporal} correlation among different time samples is negligible\footnote{This is typically the case in classic signal sampling techniques.}. As the result, the sample vectors are statistically modeled as follows: $\bx_1,\ldots,\bx_J$ are \ac{iid} such that the $n$-th time sample of the source signals are \textit{spatially} correlated. 

The spatial correlation of the time samples at $t=t_n$ is modeled via the joint probability distribution $\rmp_X\brc{x_{n}^J}$, where we define
\begin{equation}
	x_n^J\coloneqq \brc{x_{1n}, \ldots, x_{Jn}}.
\end{equation}
Hence, the joint distribution of the all signal samples is given by
\begin{equation}
	\rmp\brc{\bx^J} = \prod_{n=1}^N \rmp_X \brc{x_n^J},
\end{equation}
where the notation $\bx^J$ is defined as
\begin{equation}
	\bx^J\coloneqq \brc{\bx_{1}, \ldots, \bx_{J}}.
\end{equation}

Considering source signal $j$, an individual sensing unit collect $M_j$ linear (and potentially noisy) observations of the samples. Denoting the vector of observations by $\by_j \in \setR^{M_j \times 1}$, we can write
\begin{equation}
	\by_j=\mA_j\bx_j+\bz_j.
\end{equation}
Here, $\mA_j \in \setR^{M_j \times N}$ denotes the \textit{sensing matrix} of unit $j$ which describe the linear transform from the signal samples to the observations, and $\bz_j \in \setR^{M_j\times 1}$ is measurement noise at terminal $j$.

The observations, as well as the sensing matrices, are given to a \textit{single} data-fusion center. The data fusion center recovers the signal samples using a  \textit{joint} recovery algorithm, i.e., it finds the estimates $\bhx^J = \hat{\bx}_1,\ldots,\hat{\bx}_J$ as
\begin{equation}
	 \bhx^J= \bgg\brc{ \by_1,\ldots,\by_J \vert \mA_1, \ldots, \mA_J },
\end{equation}
via some recovery algorithm $\bgg\brc{ \cdot \vert \mA_1, \ldots, \mA_J }$. At this point, we consider a generic form for the recovery algorithms. We will later focus on a specific (but broad) class of sparse recovery algorithms which use the method of least-squares.

\subsection{Characterization of the Recovery Performance}
Before illustrating the details of the system model, let us clarify the ultimate goal of this manuscript, i.e., the asymptotic analysis of sparse recovery algorithms. To this end, we first need to define a metric which characterizes the performance of a recovery algorithm $\bgg\brc{ \cdot \vert \mA_1, \ldots, \mA_J }$. This metric is defined in the following definition:

\begin{definition}[Average distortion]
	Consider a distortion function 
	\begin{equation}
		\Delta \brc{\cdot;\cdot}: \setR^J \times \setR^J \mapsto \setR.
	\end{equation}
Using this function, the distortion between the source samples $\bx^J$ and their corresponding estimates $\bhx^J$ is determined as
	\begin{equation}
		\Delta \brc{\bhx^J; \bx^J} = \sum_{n=1}^{N} \Delta \brc{\hx_{n}^J; x_{n}^J}.
	\end{equation}
	The average distortion is then given by
	\begin{equation}
		D_N = \frac{1}{N} \Ex{	\Delta \brc{\bhx^J; \bx^J} }{}
	\end{equation}
where $\Ex{\cdot}{}$ indicates mathematical expectation.
\end{definition}

The average distortion describes the quality of the recovery algorithm. Depending on the choice of the distortion function, the average distortion determines different forms of estimation errors. For example, by setting
	\begin{equation}
	\Delta \brc{\hx_{n}^J; x_{n}^J} = \sum_{j=1}^{J} \abs{\hx_{jn} - x_{jn}}^2,
\end{equation}
the average distortion reduces to the well-known \ac{mse}.

It is important to keep in mind that the average distortion explicitly depends on the recovery algorithm. In fact, depending on the choice of $\bgg\brc{ \cdot \vert \mA_1, \ldots, \mA_J }$, the estimated samples in $\bhx^J$ change, and consequently, $D_N$ varies.

\begin{svgraybox}
Our ultimate goal is to find the average distortion for a large class of sparse recovery algorithms when the number of signal samples per each terminal, i.e., $N$, is very large. For a most known sparse recovery algorithms, this is a hard task to do, due to reasons which we explain in Section~\ref{sec:asymp}.
\end{svgraybox}

In order to address this goal, we need to specify a model for every component of the setting. We do this in the following section.

\subsection{Stochastic Model of System Components}
A typical model for the noise processes is the \ac{awgn} model. This follows from the fact that noise in sensing devices is physically caused by several random independent processes whose spectral density in the bandwidth of interest is well approximated by that of \ac{awgn}. Considering the \ac{awgn} model, the vector $\bz_j$ is considered to be an  \ac{iid} Gaussian random vector whose elements are zero-mean with variance $\sigma_j^2$.

We also model the sensing matrices as they are generated by a random process. Although stochastic modeling of noise is widely accepted, considering such model for sensing matrices requires a bit of illustration. A stochastic model for sensing matrices assumes that each sensing matrix $\mA_j$ is taken at random from a predefined ensemble. The logic behind considering such a model is as follows: From the compressive sensing literature, we know that sensing matrices require to satisfy some specific properties, such that a certain recovery performance is guaranteed~\cite{foucart13}. Many random ensembles are shown to satisfy these properties. This means that~by generating a sensing matrix from these ensembles at random, the anticipated recovery performance is achieved with a high probability. To incorporate this fact into the analysis, the classic approach is to assume that the sensing matrices are given by a random ensemble. From the mathematical viewpoint, such an assumption does not harm the generality of the analysis, as structured sensing matrices can be described by an ensemble as well.

In this manuscript, we assume that the sensing matrices are \textit{right rotationally invariant} random matrices. This is generic assumption, since it includes most well-known random ensembles, e.g., the class of \ac{iid} sensing matrices. We introduce this random ensemble in the sequel. However, before defining that let us first define the density of states for a given matrix.
\begin{definition}[Density of states]
	Let $\mS \in \setR^{N \times N}$ be a self-adjoint square matrix whose eigenvalues are given by $\lambda_1, \ldots, \lambda_N \in \setR$. The density of states for this matrix is defined as the empirical \ac{cdf} of its eigenvalues, i.e., for $\lambda\in \setR$
\begin{equation}
	F_{\mS}^N \brc{\lambda} = \frac{1}{N} \sum_{n=1}^N \mone \set{ \lambda_n < \lambda }.
\end{equation}
\end{definition}

We are now ready to define right rotationally invariant random matrices. 

\begin{definition}[Right rotationally invariant matrices]
$\mA_j \in \setR^{M_j \times N}$ is right rotationally invariant if the Gramian $\mJ_j=\mA_j^{\trp} \mA_j$ has the decomposition 
\begin{equation}
	\mJ_j= \mU_j \mD_j \mU_j^{\trp}
\end{equation}
where $\mU_j$ and $\mD_j$ have the following properties:
\begin{enumerate}
	\item  $\mU_j\in\setR^{N\times N}$ is a Haar distributed matrix\footnote{A Haar distributed matrix is a random matrix generated uniformly on the set of all orthogonal matrices.}.
	\item $\mD_j\in\setR^{N\times N}$ denotes the diagonal matrix of eigenvalues whose density of states is equal to the density of states of $\mJ_j$, i.e., $F_{\mD_j}^N \brc{\lambda}  = F_{\mJ_j}^N \brc{\lambda} = F_{j}^N \brc{\lambda}$.
\end{enumerate}
\end{definition}

The class of right rotationally invariant matrices includes most well-known random ensembles in compressive sensing. Examples of such ensembles are \ac{iid} and row-orthogonal random sensing matrices. Note that different forms of random matrices will have different densities of states. We clarify this point further in Section~\ref{sec:asymp} when we formally formulate the asymptotic analysis of a sparse recovery algorithm.

\subsection{Stochastic Model for Jointly Sparse Signals}
\label{sec:joint_sp}
As indicated, we assume that there exists a spatial correlation among the signals for various sources. Noting that the signals are sparse, we interpret this spatial coupling as \textit{joint sparsity}. From statistical viewpoint, this means that the joint \ac{pdf} $\rmp_X\brc{x^J_n}$ contains one or multiple Dirac impulses. These impulses occur on subspaces of $\setX^J$, where at least one of the vector entries is set to zero. 

To give an intuition on joint sparsity, we focus in the sequel on a special joint sparsity model in which the sample $n$ of signal $j$ is written as
\begin{equation}
	x_{jn}= c_{n} w_{0n} + s_{ 0 n } w_{jn} + s_{j n} u_{jn}.
\end{equation}
Here, $s_{0n}$, $w_{0n}$, $c_{ n }$, $w_{jn}$, $s_{j n}$ and $u_{jn}$ are independent \ac{iid} sequences whose distributions are as follows:
\begin{enumerate}
	\item $w_{0n}$, $w_{jn}$ and $u_{jn}$ are random variables on $\setX$ whose probabilities of being zero are equal to zero.
	\item $s_{0n}$, $c_{ n }$ and $s_{j n}$ are Bernoulli random variables with 
\begin{subequations}
		\begin{eqnarray}
		\Pr\set{c_{n}=1} &=& 1-\Pr\set{c_{n}=0}=\mu_{\rmc} \\
		\Pr\set{s_{0n}=1} &=& 1-\Pr\set{s_{0n}=0}=\mu_0\\
		\Pr\set{s_{jn}=1} &=& 1-\Pr\set{s_{jn}=0}=\mu_j.
	\end{eqnarray}
\end{subequations}
\end{enumerate}

In this model, the samples of a terminal are given as the superposition of three sparse components. The first component, whose $n$-th entry is given by $ c_{n} w_{0n} $, is a sparse vector which is common among the all terminals. The second sparse component, represented by $s_{ 0 n } w_{jn}$ for $n\in\dbc{N}$, has a common support\footnote{By support, we refer to the index of non-zero entries in a vector.} across all the terminals; however, the value of the non-zero entries are drawn from independent processes. The last component contains a sparse signal whose support and non-zero entries are independently generated for each terminal.

Although the given model for joint sparsity is not the most general one, it includes most well-known sparse recovery settings in the literature. In the sequel, we address the main settings for sparse recovery. These settings are derived from our system model as special cases.

\subsection{Special Cases}
The three main settings for sparse recovery are classical compressive sensing, the problem of \ac{mmv}, and \ac{dcs}. In the sequel, we briefly go through these settings and illustrate how they are derived from our generic multi-terminal setting.

\subsubsection{Classical Compressive Sensing}
In classical compressive sensing, also called the single measurement vector problem, a sparse signal is observed linearly via a \textit{single} terminal and is to be recovered from the underdetermined set of observations. This settings is simply derived by setting $J=1$ in our model. As spatial correlation among terminals has no meaning in this case, one can further set $c_n = s_{0 n} = 0$ for $n\in \dbc{N}$ in the sparsity model. 

\subsubsection{Multiple Measurement Vectors}
In \ac{mmv}, multiple sparse signals are observed with a common sensing matrix and recovered at a single data fusion center. This setting is straightforwardly derived from our model by letting
\begin{equation}
	\mA_1 = \ldots = \mA_J.
\end{equation}

In general, the joint sparsity model given in Section~\ref{sec:joint_sp} is a valid model in \ac{mmv}. Nevertheless, in many applications of \ac{mmv}, it is common to assume a  the \textit{common support} model for the spatial correlation. This model assumes that the samples of different terminals have common support; however, the non-zero entries are drawn from independent processes. The common support model is derived from the joint sparsity model in Section~\ref{sec:joint_sp} by setting $s_{jn}=c_n = 0$ for $n\in\dbc{N}$ and $j\in\dbc{J}$.

\subsubsection{Distributed Compressive Sensing}
\label{sec:dcs}
\ac{dcs} describes the most generic setting which fits to our model. In this problem, the jointly sparse signals of different terminals are observed with different matrices and recovered at a common fusion center. Similar to \ac{mmv}, the joint sparsity model in Section~\ref{sec:joint_sp} is generally valid for \ac{dcs}. A common model is however the \textit{common-innovation} model in which the signal of each terminal is given as a common sparse component superposed by an independent sparse  innovation term. This model is derived from the one given in Section~\ref{sec:joint_sp} by setting $s_{jn} = 0$ for $n\in\dbc{N}$ and $j\in\dbc{J}$.

\section{RLS-Based Recovery Algorithms}
We focus on the class of \ac{rls}-based recovery algorithms. These algorithms recover the signal samples by minimizing a penalized residual sum of squares. In general, an \ac{rls}-based algorithm is of the following form:
\begin{equation}
	\bgg\brc{ \by_1,\ldots,\by_J \vert \mA_1, \ldots, \mA_J } = \argmin_{\bv_1, \ldots,\bv_J \in \setX^N}  \sum_{j=1}^J \frac{1}{2\lambda_j} \norm{\by_j-\mA_j\bv_j}^2 + u_{\rmv} \brc{\bv^J}. \label{eq:RLS}
\end{equation}
Here, $\bv^J \coloneqq \brc{\bv_1, \ldots, \bv_J}$, and $u_\rmv\brc{\cdot} : \setR^{JN\times 1} \mapsto \setR^+$ is the \textit{regularization function} which penalizes the residual sum of squares, and $\norm{\cdot}$ denotes the Euclidean norm. In the sequel, we assume that $u_\rmv\brc{\cdot} $ decouples, i.e., there exist $u\brc{\cdot} : \setR^{1\times J} \mapsto \setR^+$, such that
\begin{equation}
	u_{\rmv}(\bv^J)=\sum_{n=1}^N u\brc{v^J_{n}}.
\end{equation}
Furthermore, $\lambda_1, \ldots, \lambda_J$ are tunable factors, referred to as \textit{regularizers}. 

The interest in this class of recovery schemes comes from its generality. In fact, the recovery scheme in \eqref{eq:RLS} includes a diverse set of sparse recovery algorithms, e.g., $\ell_p$-norm minimization techniques in classical compressive sensing and $\ell_{p,q}$-norm minimization techniques in \ac{dcs}; we will discuss these examples with more details later on in this section. But before that, let us give a Bayesian interpretation of an \ac{rls}-based recovery scheme.

\subsection{Bayesian Interpretation}
In the Bayesian view point, an \ac{rls}-based recovery scheme is a \ac{map} estimator in which
\begin{enumerate}
	\item The \textit{prior} joint distribution of samples at $t_n$ is \textit{postulated} to be proportional to $\exp\set{- u\brc{\cdot}} $. This means that $\rmp_X\brc{x_n^J}$ is \textit{assumed} to be
	\begin{equation}
		\rmp_X\brc{x_n^J} = \frac{\exp\set{- u\brc{\cdot}} }{Z}
	\end{equation}
for some normalization factor $Z$.
\item  The noise processes are postulated to be Gaussian.
	\item The variance of noise at terminal $j$ is \textit{postulated} to be $\lambda_j$.
\end{enumerate}
It is worth mentioning that the postulated parameters are not necessarily matched to the true ones. Hence, the \ac{map} estimator is in general \textit{mismatched}.

\subsection{Special Cases}
As mentioned, the \ac{rls}-based scheme includes several sparse recovery algorithms. In the sequel, we discuss two well-known examples, namely $\ell_p$-norm minimization for classical compressive sensing and $\ell_{p,q}$-norm minimization for \ac{dcs}.

\subsubsection{$\ell_p$-Norm Minimization}
Most algorithms in compressive sensing with a single terminal recover the sparse signal by finding a vector of samples whose residual sum of squared is bounded, i.e., finding $\bv$, such that\footnote{Note that index $j$ is dropped, as we consider a single terminal setting.}
\begin{equation}
\norm{\by -\mA \bv}^2 \leq \epsilon
\end{equation}
for some $\epsilon$, and whose $\ell_p$-norm for some $p$ is minimum. The most common choice of $p$ is $p=1$ which results in the \ac{lasso} or basis pursuit algorithm. 

Using the method of Lagrange multipliers, it is shown that there exists a regularizer $\lambda$, for which the \ac{rls}-based algorithm with decoupled regularization function $u\brc{v_n} = \abs{v_n}^p$ performs identical to this algorithm.

\subsubsection{$\ell_{p,q}$-Norm Minimization}
For multi-terminal settings, the classic $\ell_p$-norm minimization techniques are often extended to $\ell_{p,q}$-norm minimization techniques. The feasible set in this case is constructed with the same approach, i.e., finding $\bv_1,\ldots,\bv_J$ for which
\begin{equation}
	\norm{\by_j -\mA_j \bv_j}^2 \leq \epsilon_j
\end{equation}
with some $\epsilon_j$ for $j\in\dbc{J}$. The recovered sampled are then found by searching the feasible for vectors whose $\ell_{p,q}$-norm is minimum. For a collection of $J$ vectors $\bv_1,\ldots,\bv_J$, the $\ell_{p,q}$-norm is defined as
\begin{equation}
\norm{\bv_1, \ldots, \bv_J}_{p,q} = \sum_{n=1}^N \left[ \brc{\sum_{j=1}^J \abs{v_{jn}}^p}^{1/p} \right]^q .
\end{equation}
The most well known $\ell_{p,q}$-norm minimization technique is the group \ac{lasso} technique in which $p=2$ and $q=1$.

Similar to $\ell_{p}$-norm minimization, one can invoke the method of Lagrange multipliers and show that there exist regularizers $\lambda_1,\ldots,\lambda_J$, for which the \ac{rls}-based algorithm with decoupled regularization function $u\brc{v_n^J} = \brc{\norm{v_n^J}_p}^q$ performs identical to $\ell_{p,q}$-norm minimization.

\section{Asymptotic Characterization}
\label{sec:asymp}
Now that the system model and recovery algorithm are presented, we are ready to formally formulate the asymptotic performance of a sparse recovery algorithm. For asymptotic analysis, we consider a sequence of settings. The number of signal samples $N$ and the number of measurements at terminal $j$, i.e., $M_j$, in this sequence grow large, such that $M_j$ is a deterministic function of $N$. We assume that $N$ grows unboundedly large, and $M_j$ grows with $N$ linearly. This means that there exists a fixed $\rho_j$ (typically $\rho_j\leq 1$) for each $j\in\dbc{J}$, such that
\begin{align}
	\rho_j \coloneqq \lim_{N \uparrow \infty} \frac{M_j}{N} < \infty.
\end{align}
We refer to $\rho_j$ as the $j$-th terminal compression ratio.

For every \ac{dsn} in the sequence, we use an \ac{rls}-based algorithm to recover the signal samples. Let $D_N$ denote the average distortion between the true signal samples and their estimates in the \ac{dsn} whose index is $N$. The asymptotic analysis intends to find the asymptotic limit of this sequence distortions, i.e., 
\begin{align}
	D \coloneqq \lim_{N \uparrow \infty} D_N.
\end{align}

The derivation of $D$ and its dependence on the various model parameters are not straightforward from both analytical and computational points of view. In fact, depending on the regularization function, the derivation of $D$ deals with one or two of the following issues:
\begin{itemize}
	\item For some regularization functions, the \ac{rls}-based algorithm solves a \textit{convex} optimization problem. This means that the recovery is posed as a linear programming. Hence, it is performed in polynomial time. Although \ac{rls}-based recovery in this case is computationally tractable, there is no guarantee that the problem is also \textit{analytically} tractable. For asymptotic analysis, one needs to determine the sequence of average distortions for any integer index $N$ and take the limit when $N$ goes to $\infty$. For some particular \ac{rls}-based algorithms, this task can be done via basic analytic tools; nevertheless, there are several forms whose limit is not known analytically via the basic tools.
	\item Several \ac{rls}-based algorithms are not only analytically, but also computationally intractable. An example is the $\ell_0$-norm minimization algorithm in which the regularization function is proportional to the $\ell_0$-norm. In such cases, the recovery algorithm deals with an \ac{np}-hard problem \cite{foucart13}. Clearly, for these forms, asymptotic characterization is not tractable.
\end{itemize}

The above analytical and computationally issues can be addressed via the replica method. As it becomes clear later, the replica method invokes several non-rigorous tricks to bypass the analytical obstacles of the problem. The term \textit{non-rigorous tricks} will be clarified in the next sections of this manuscript while we illustrate how the replica method exactly does that. 

Now that the asymptotic analysis is formulated in principle, we can state explicitly our main purpose as follows:
\begin{svgraybox}
The main purpose of this manuscript is to illustrate how the asymptotic distortion $D$ is derived for an \ac{rls}-based algorithm via the replica method.
\end{svgraybox}

\subsection{Stieltjes and $\rmR$-Transforms}
Before we start with the illustration of the replica method, we give some basic definitions which are used throughout the derivations via the replica method. These definitions enables us to represent compactly the statistics of the sensing matrices.

To start with the definitions, consider the sequence of \acp{dsn} indexed by $N$. For each terminal, there exists a corresponding sequence of densities $F_j^N\brc{\lambda}$ which for a given index $N$ describes the density of states of $\mJ_j = \mA_j^\trp \mA_j$. We assume that this sequence converges as $N\rightarrow\infty$ to a deterministic density of state $F_j\brc{\lambda}$ for each $j\in \dbc{J}$. For these asymptotic densities, the Stieltjes and $\rmR$-transforms are defined as follows \cite{tulino2004random}:

\begin{definition}[Stieltjes Transform]
	For the asymptotic density $F_j \brc{\lambda}$, the Stieltjes transform is given by 
	\begin{equation}
		\rmG_j\brc{s}= \int_{-\infty}^{+\infty} \frac{1}{\lambda-s} \dif F_j \brc{\lambda}
	\end{equation}
	for some complex $s$ with $\Im{s} \geq 0$, where $\Im{s}$ is the imaginary part of $s$.
\end{definition}

\begin{definition}[$\rmR$-Transform]
		For the asymptotic density $F_j \brc{\lambda}$, the $\rmR$-transform is defined as
		\begin{equation}
			\rmR_j \brc{\omega} = \rmG_j^{-1} \brc{-\omega} - \frac{1}{\omega}
		\end{equation}
	such that 
	\begin{equation}
		\lim_{\omega\downarrow 0} \rmR_j \brc{\omega} = \int_{-\infty}^{+\infty} \lambda \dif F_j \brc{\lambda}.
	\end{equation}
	$\rmG_j^{-1} \brc{\cdot}$ denotes the inverse of the Stieltjes transform with respect to composition.
\end{definition}

This definition of the $\rmR$-transform is further extended to matrix arguments: Consider a matrix $\mS_{N \times N}$ with the eigendecomposition
\begin{equation}
	\mS=\mathbf{\Sigma}\mathbf{\Lambda} \mathbf{\Sigma}^{-1}.
\end{equation}
For this matrix, we use the notation $\mathrm{R}_j \brc{\mS}$ to refer to
\begin{equation}
\mathrm{R}_j \brc{\mS} \coloneqq \mathbf{\Sigma} \ \mathrm{Diag}\set{ \mathrm{R}_j\brc{\lambda_1}, \ldots, \mathrm{R}_j \brc{\lambda_n}} \ \mathbf{\Sigma}^{-1}
\end{equation}
where $\mathrm{Diag}\set{ a_1,\ldots,a_N}$ denotes an $N\times N$ diagonal matrix whose diagonal entries are $a_1,\ldots,a_N$.

\section{Building a Bridge to Statistical Mechanics}
As mentioned before, the replica method was initially developed in statistical mechanics for analysis of spin glasses. Nevertheless, it found its way to several other fields, such as coding, information theory and signal processing. The key point in employing the replica method for asymptotic analysis is to make a connection between the problem in hand and the theory of spin glasses. In this section, we illustrate how this connections is made. To this end, we need first to give a quick overview on basic definitions in statistical mechanics.

\subsection{Introduction to Statistical Mechanics}
A thermodynamic system consists of $N$ particles with each having a microscopic parameter $v_n\in \setV $ for $n\in[N]$ and some set $\setV$. This parameter describes a macroscopic property of the corresponding particle, e.g., the velocity. In general, a microscopic parameter could be a vector of continuous or discrete entries. For sake of brevity, we assume that $v_n$ is a \textit{continuous} scalar. The extension to cases with discrete $\setV$ can be followed in \cite[Chapter 3]{bereyhi2020thesis}. For this system, the \textit{microstate} is defined as a vector in $\setV^{N }$ which collects microscopic parameters of all the particles, i.e.
	\begin{equation}
		\bv= \left[ v_1,\ldots,v_N\right]^\trp.
	\end{equation}

Corresponding to this system, a \textit{Hamiltonian} $\mae\brc{\cdot}$ is defined which describes the physical properties of the system. The Hamiltonian is a function which assigns to microstate $\bv$ a non-negative energy level $\mae\brc{\bv}$.

\begin{trailer}{A Side Note}
Here, we have defined the Hamiltonian in an abstract form. For a physical system, the explicit form of the Hamiltonian is derived from the physical theories which describe the interactions of microscopic parameters in the system.
\end{trailer}

For a thermodynamic system, the explicit calculation of macroscopic parameters is intractable\footnote{This follows the same reasons given in Section~\ref{sec:asymp} for the asymptotic analysis of \ac{rls}-based algorithms.}. To address this issue, statistical mechanics follows a stochastic approach. In this approach, the microstate is considered to be a random vector whose distribution depends on the \textit{temperature}. We denote this distribution by $\rmp_{\beta}\brc{\bv}$ where $\beta$ is the \textit{inverse temperature}, i.e., $\beta = 1 / T$ with $T$ being the temperature.

Using stochastic analysis, statistical mechanics derives physical features of the thermodynamic system from this stochastic model. These physical features are known as \textit{macroscopic} parameters of the system. Mathematically, a thermodynamic system can be described by two main macroscopic parameters, namely the \textit{entropy} and \textit{free energy}\footnote{In fact, the main two macroscopic parameters of a thermodynamic system are entropy and energy. The free energy is derived by applying the second law of thermodynamic as the Lagrange dual function. We however use directly the free energy in our formulation, for sake of brevity.}. These parameters are defined as follows:

\begin{definition}[Normalized entropy]
	For a given thermodynamic system with $N$ particles, the normalized entropy at inverse temperature $\beta$ is defined as
	\begin{align}
		\maH_N \brc{\beta} \coloneqq -\frac{1}{N} \int\limits_{\setV^{N }} \rmp_{\beta}\brc{\bv} \log \rmp_{\beta}\brc{\bv} \dif \bv
	\end{align}
		\label{def:entropy}
\end{definition}

\begin{definition}[Normalized free energy]
	Consider a thermodynamic system with $N$ particles and Hamiltonian $\mae\brc{\cdot}$. At inverse temperature $\beta$, the normalized free energy is defined as
	\begin{align}
		\maF_N \brc{\beta} \coloneqq \frac{1}{N} \Ex{\mae(\bv)}{ } -\frac{1}{\beta} \maH_N \brc{\beta}. \label{eq:int-3}
	\end{align}
		\label{def:free_energy}
\end{definition}

\subsubsection{Second Law of Thermodynamics}
\label{sec:2ndLaw}
The fundamental rule in stochastic analysis of thermodynamic systems is the second law of thermodynamic. This law indicates that the microstate in thermal equilibrium\footnote{This means that there is no energy flow.} is distributed such that the free energy is minimized. Since $\maF_N \brc{\beta}$ is convex with respect to $\rmp_{\beta} \brc{\bv}$, it is concluded that the microstate is distributed with the \textit{Boltzmann-Gibbs} distribution. This means that at thermal equilibrium
\begin{equation}
	\rmp_{\beta} \brc{\bv}= \frac{ \exp\set{-\beta \mae\brc{\bv} } }{ \maz_N\brc{\beta} } 
\end{equation}
at inverse temperature $\beta$. $\maz_N\brc{\beta}$ in the denominator is a normalization factor, i.e., 
\begin{equation}
	\maz_N\brc{\beta}= \int\limits_{\setV^{N }} \exp\set{-\beta \mae\brc{\bv} } \dif \bv ,
\end{equation}
and is called the \textit{partition function}. $\rmp_{\beta} \brc{\bv}$ reduces to some well-known distributions for several choices of the Hamiltonian, e.g., it reduces to the Gaussian distribution when $\mae\brc{\bv} \propto \norm{\bv}^2$.

\newpage
\begin{trailer}{A Side Note}
The stated form of the second law of thermodynamics is a simplified interpretation of the original form. In fact, the law states that the entropy in an isolated system grows constantly. This is interpreted as a constrained optimization problem in which the normalized entropy is maximized subject to an energy constraint. Using the method of Lagrange multipliers, the free energy is derived as the objective function of the dual unconstrained optimization. It is then shown that the Lagrange multiplier is in fact the temperature.
\end{trailer}

Substituting the Boltzmann-Gibbs distribution in the definition of the free energy, it is concluded that
\begin{equation}
	\maF_N \brc{\beta} = -\frac{1}{\beta N} \log \maz_N\brc{\beta}. \label{eq:Free}
\end{equation}
\begin{svgraybox}
	This is a fundamental identity indicating that the free energy of a system in thermal equilibrium is calculated explicitly from the partition function. Starting from this equation, it is shown that all other macroscopic parameters of the system are directly derived from  $\maF_N \brc{\beta}$. For instance, 
	\begin{equation}
		\maH_N \brc{\beta}  = \beta^2  \frac{\dif}{\dif \beta} \maF_N \brc{\beta}. \label{eq:Entropy_Free1}
	\end{equation}
Therefore, the partition function completely describes the macroscopic features of the system in thermal equilibrium.
\end{svgraybox}

\subsubsection{Spin Glasses}
\textit{Spin glasses} are thermodynamic systems whose particles choose to interact randomly. This means that the Hamiltonian of a spin glass is not only a function of the microstate, but also a randomizer. This randomizer is realized once from a random ensemble and remains fixed as the system is in thermal equilibrium\footnote{In statistical mechanics, this randomizer is known to have \textit{quenched} randomness. This is different from the type of randomness considered for the microstate which is called \textit{annealed} randomness. }. 

Similar to thermodynamic systems, the stochastic analysis of spin glasses follows the second law of thermodynamic. Let $\mOmega$ denote the randomizer of a spin glass. The Hamiltonian of this spin glass is given by
\begin{equation}
	\mae \brc{\cdot\vert \mOmega} : \setV^{N } \mapsto \setR^+.
\end{equation}
In other words, for every realization of $\mOmega$, we have a specific Hamiltonian function. By the same lines of derivations explained in Section~\ref{sec:2ndLaw}, one can show that \textit{conditioned} to the randomizer, the microstate of the spin glass in thermal equilibrium is distributed with the Boltzmann-Gibbs distribution. This means that
\begin{equation}
	\rmp_{\beta} \brc{\bv\vert \mOmega}= \frac{ \exp\set{-\beta \mae\brc{\bv\vert \mOmega} } }{ \maz_N\brc{\beta\vert \mOmega} }
\end{equation}
with random partition function  
\begin{equation}
	\maz_N\brc{\beta\vert \mOmega}= \int\limits_{\setV^{N }} \exp\set{-\beta \mae\brc{\bv\vert \mOmega} } \dif \bv
\end{equation}
The normalized free energy in thermal equilibrium is hence written as
\begin{equation}
	\maF_N \brc{\beta\vert \mOmega} = -\frac{1}{\beta N} \log \maz_N\brc{\beta\vert \mOmega},
\end{equation}
and the \textit{conditional} entropy is determined from the free energy by
\begin{equation}
	\maH_N \brc{\beta\vert \mOmega}  = \beta^2  \frac{\dif}{\dif \beta} \maF_N \brc{\beta\vert \mOmega}.  \label{eq:Entropy_Free2}
\end{equation}
In the remaining, we focus on spin glasses. This is due to the fact that our problem is formulated in terms of a spin glass.

\subsubsection{Thermodynamic Limit}
Spin glasses are studies in the \textit{thermodynamic limit}. This means that the macroscopic parameters are derived for the case, in which the number of particles tends to infinity, i.e., the asymptotic limit $N\uparrow \infty$. Suggested by physical intuition, in the thermodynamic limit, a spin glass has deterministic macroscopic parameters. This means that in the asymptotic limit, the free energy $\maF_N \brc{\beta\vert \mOmega}$ converges to its expected value\footnote{Here, the expectation is taken over the randomizer $\mOmega$.}. This property of spin glasses is known as \textit{self averaging}; more discussions in this respect can be followed in \cite{pastur1991absence,guerra2002thermodynamic,guerra2002infinite}. 

Following the self-averaging property, the free energy of a spin glasses in the thermodynamic limit is calculated as follows:
\begin{enumerate}
	\item Determining sequence of \textit{expected} free energies $\bar{\maF}_N\brc{\beta\vert\mOmega}$ indexed by$N$ as
		\begin{equation}
		\bar{\maF}_N \brc{\beta} =  \Ex{\maF_N\brc{\beta\vert\mOmega}}{ }, \label{eq:ExpectedFreeEnergy}
	\end{equation}
where the expectation is taken with respect to $\mOmega$. 
	\item Taking the asymptotic limit of the expected sequence, i.e., calculating
	\begin{equation}
		\bar{\maF} \brc{\beta} = \lim_{N\uparrow \infty} \bar{\maF}_N \brc{\beta}.
	\end{equation}
\end{enumerate}

\subsubsection{Averaging Trick}
Before we start with the derivations, let us illustrate the key \textit{averaging trick} in statistical mechanics. Consider a function $\psi_N\brc{\cdot}$ which for each microstate $\bv\in\setV^N$ determines an scalar parameter. The macroscopic parameter corresponding to this function is defined as
\begin{equation}
	\bar{\psi}_N = \frac{1}{N} \Ex{\psi_N\brc{\bv} }{ }
\end{equation}
where the expectation is taken first with respect to the conditional Boltzmann-Gibbs distribution, i.e., $\rmp_{\beta} \brc{\bv\vert \mOmega}$, and then with respect to $\mOmega$.

The classic approach for determining $\psi_N$ in statistical mechanics is to use the averaging trick. This trick modifies the partition function with a dummy factor $h$ as follows
\begin{equation}
	\maz_N \brc{\beta,h|\mOmega}= \int\limits_{\bv\in \setV^{N } } \exp\set{-\beta \mae\brc{\bv\vert \mOmega} + h \psi_N\brc{\bv} } \dif \bv
\end{equation}
For this modified partition function, the normalized free energy, conditioned to a realization of the randomizer,is
\begin{equation}
	\maF_N \brc{\beta,h\vert \mOmega} = -\frac{1}{\beta N} \log \maz_N\brc{\beta,h\vert \mOmega},
\end{equation}
and its expected value $\bar{\maF}_N \brc{\beta,h}$ is determined by calculating the expectation over $\mOmega$, i.e., as in \eqref{eq:ExpectedFreeEnergy}.

By standard derivations, its is readily shown that
	\begin{equation}
		\bar{\psi}_N = -\beta \frac{\partial}{\partial h} \bar{\maF}_N \brc{\beta,h}\vert_{h=0}
	\end{equation}
Exchanging limiting procedures, one has in the thermodynamic limit
\begin{subequations}
	\begin{eqnarray}
	\bar{\psi} &\coloneqq& \lim_{N \uparrow \infty} \bar{\psi}_N\\
	 &=& -\beta \frac{\partial}{\partial h} \bar{\maF} \brc{\beta,h}\vert_{h=0}.
\end{eqnarray}
\end{subequations}

\subsection{Corresponding Spin Glass}
The connection between the sparse recovery problem and statistical mechanics is illustrated by introducing the concept of \textit{corresponding spin glass}. In fact, for an \ac{rls}-based recovery algorithm, we can define an \textit{imaginary} spin glass whose macroscopic parameters are the asymptotic performance metric of the recovery algorithm. We clarify this connection in the sequel.

Remember the system model in Section~\ref{sec:SysMod} with sensing matrices $\mA_1, \ldots \mA_J$ and observation vectors $\by_1,\ldots,\by_J$. We define corresponding spin glass as follows:

\begin{trailer}{Corresponding Spin Glass}
The corresponding spin glass is a spin glass whose microstate is described by $\bv^J = \brc{\bv_1, \ldots, \bv_J}$ where $\bv_j \in \setX^{N}$ for $j\in \dbc{J}$. The randomizer of this spin glass is
\begin{equation}
	\mOmega = \set{\mA_1, \ldots, \mA_J, \by_1, \ldots, \by_J},
\end{equation}
and its Hamiltonian is
\begin{equation}
	\mae\brc{\bv^J \vert \mOmega }=	\sum_{j=1}^J \frac{1}{2\lambda_j} \norm{\by_j-\mA_j\bv_j}^2 + u_{\rmv} \brc{\bv^J}. \label{eq:Hamiltonian_SpinGlass}
\end{equation}
\end{trailer}

From our earlier discussions, we know that at inverse temperature $\beta$, the microstate in thermal equilibrium is conditionally distributed with
\begin{equation}
	\rmp_{\beta} \brc{\bv^J \vert \mOmega}= \frac{\exp\set{-\beta \mae\brc{\bv^J\vert \mOmega}}}{\maz_N \brc{\beta\vert \mOmega}} 
\end{equation}
where partition function $\maz_N \brc{\beta|\by,\mA}$ reads
\begin{equation}
	\maz_N \brc{\beta\vert \mOmega} = \int\limits_{\bv_j\in \setX^{N }} \exp\set{-\beta \mae\brc{\bv^J\vert \mOmega} } \dif \bv^J.
\end{equation}

The key property of this spin glass which connects it to our sparse recovery problem is its \textit{ground state property}. 

\begin{trailer}{Ground State Property}
For a given realization of $\mOmega$, assume that the Hamiltonian has a unique minimizer denoted by $\bv^J_{\star}\brc{\mOmega}$. Then, as the temperature goes to zero, i.e., $\beta \uparrow\infty$, the microstate of the corresponding spin glass converges in distribution to the deterministic vector $\bv^J_\star\brc{\mOmega}$. This means that for every realization of $\mOmega$
\begin{equation}
	\lim_{\beta\uparrow\infty} \rmp_{\beta} \brc{\bv^J\vert \mOmega} 		
	= \begin{cases} 
		1 &\bv^J=\bv^J_\star\brc{\mOmega}\\
		0 & \bv^J \neq \bv^J_\star\brc{\mOmega}
	\end{cases}. \label{eq:GroundState}
\end{equation}
\end{trailer}

In fact, this is a well-known property in statistical mechanics: At zero temperature, the microstate converges in distribution to a realization whose energy level is minimized. The appellation follows the fact that this realization, i.e., $\bv^J_{\star}\brc{\mOmega}$, is called the \textit{ground state} of the system.

\begin{svgraybox}
	The ground state property clarifies the connection between our problem and this spin glass: In fact, the ground state is what the \ac{rls}-based algorithm recovers, i.e.,
\begin{equation}
	\bv^J_{\star}\brc{\mOmega} = \bhx^J.
\end{equation}
In other words, as the temperature goes to zero the microstate of the corresponding spin glass converges to the signal samples which are recovered via the algorithm. Hence, the performance metrics of this sparse recovery algorithm, e.g., the asymptotic distortion, are derived as the macroscopic parameters of this spin glass at zero temperature.
\end{svgraybox}

The corresponding spin glass shows several other interesting properties. Interested readers are suggested to see \cite[Chapter 3]{bereyhi2020thesis}.

\subsubsection{Asymptotic Distortion as a Macroscopic Parameter}
The main purpose of this manuscript is to determine the asymptotic distortion. As indicated, this metric can be defined as a macroscopic parameter of the corresponding spin glass. To show that, consider the following macroscopic parameter
\begin{equation}
	\psi_N \brc{\bv^J} =  \Delta \brc{\bv^J; \bx^J} 
\end{equation}
where $\bx^J$ refer to the true signal samples. The macroscopic parameter defined by this function is
\begin{subequations}
	\begin{eqnarray}
	\bar{\psi} &=& \lim_{N \uparrow \infty} \frac{1}{N} \Ex{\psi_N\brc{\bv} }{ }\\
	&=& \lim_{N \uparrow \infty} \frac{1}{N} \Ex{ \Delta \brc{\bv^J; \bx^J}  }{ }
\end{eqnarray}
\end{subequations}
As the temperature goes to zero, $\beta \uparrow$, the microstate $\bv^J$ converges to $\bhx^J$. Hence, at zero-temperature we have
\begin{subequations}
	\begin{eqnarray}
	\bar{\psi} &\to&  \lim_{N \uparrow \infty} \frac{1}{N} \Ex{ \Delta \brc{\bhx^J; \bx^J}  }{ }\\
	&=& \lim_{N \uparrow \infty} D_N \\
	&=& D.
\end{eqnarray}
\end{subequations}
The last equation clarifies how the asymptotic distortion is derived from the corresponding spin glass.

\begin{svgraybox}
	Using the averaging trick, we can find $D$ from the following expected modified free energy in the thermodynamic limit
\begin{equation}
	\bar{\maF }\brc{\beta,h} = - \lim_{N \uparrow \infty} \frac{1}{\beta N}  \Ex{\log\maz_N\brc{\beta,h\vert \mOmega}}{}, \label{eq:FinalFreeEnergy}
\end{equation}
at zero temperature as
\begin{equation}
	D = - \lim_{\beta\uparrow \infty} \beta \frac{\partial}{\partial h} \bar{\maF} \brc{\beta,h}\vert_{h=0},
\end{equation}
where the partition function is given by
\begin{equation}
	\maz_N \brc{\beta,h|\mOmega}= \int\limits_{\bv_j\in \setX^{N } } \exp\set{-\beta \mae\brc{\bv^J\vert \mOmega} + h \Delta \brc{\bv^J; \bx^J}  } \dif \bv^J.
\end{equation}
\end{svgraybox}

\subsection{The Replica Method}
The variational problem derived in terms of the corresponding spin glass suffers from the same intractability issue we observed in the original problem. In the original problem, the intractability issue was due to the optimization problem. This is now transformed to a \textit{logarithmic expectation} in \eqref{eq:FinalFreeEnergy}. This is not a trivial task, and hence keeps the problem still intractable.

The replica method tries to calculate this logarithmic expectation with a series of tricks. The first step is to use the Riesz identity \cite{riesz1930valeurs}:

\begin{trailer}{Riesz Identity}
	For a non-negative random variable $X$, we have
	\begin{equation}
		\Ex{\log X}{} = \lim_{\theta \downarrow 0} \frac{\log \Ex{X^\theta}{}}{\theta}
	\end{equation}
\end{trailer}  

Using this identity, one can rewrite the logarithmic expectation of \eqref{eq:FinalFreeEnergy} as 
\begin{equation}
  \Ex{\log\maz_N\brc{\beta,h\vert \mOmega}}{} = \lim_{\theta \downarrow 0} \frac{ \log \Ex{\maz_N^\theta \brc{\beta,h\vert \mOmega}}{}  }{ \theta }.
\end{equation}
The right hand side now deals with logarithm of an expectation. The problem is however still intractable, since $\theta$ in the right hand side of the identity is a \textit{real} scalar: The intractability of logarithmic expectation is now transformed to the challenge of \textit{calculating real moments}. Here, the second step is applied:

\begin{trailer}{Replica Continuity}
We assume that the moment function is real analytic on the real axis. This means that the function
\begin{equation}
	f_{\rm M}\brc{\theta} =   \Ex{\maz_N^\theta\brc{\beta,h\vert \mOmega}}{} 
\end{equation}
can be analytically continued from integer values to all real $\theta$.
\end{trailer}

\begin{svgraybox}
	This second step is \textit{not mathematically rigorous}. This is why the replica method is often called the \textit{replica trick}. The available results suggest that this is a valid assumption; however, the proof is still an open problem.
\end{svgraybox}

Assuming $\theta$ to be an integer finally resolves the intractability issue at the expense of loosing mathematical rigor. We now can write the moment function as\footnote{In the notation, we drop the integration set for sake of compactness.}
\begin{subequations}
\begin{eqnarray}
	f_{\rm M}\brc{\theta} &=&   \Ex{\maz_N^\theta\brc{\beta,h\vert \mOmega}}{} \\
	&=& \Ex{\prod_{a=1}^\theta 
\int \exp\set{-\beta \mae\brc{\bv_a^J\vert \mOmega} + h \Delta \brc{\bv_a^J; \bx^J}  } \dif \bv_a^J	
}{}\\
&=& 	\int \Ex{\exp\set{-\beta \sum_{a=1}^\theta \mae\brc{\bv_a^J\vert \mOmega} + h \Delta \brc{\bv_a^J; \bx^J}  }}{} \dif \bv_1^J \ldots \dif \bv_\theta^J. \label{eq:FinalIntegral}
\end{eqnarray}
\end{subequations}

The latter integral is complicated but tractable. The main remained task is to calculate this integral and find it as an analytic function in $\theta$. We then replace it into the Riesz identity, and take the limits. In the sequel, we give a quick overview on the derivations.

\section{The Replica Analyses}
The derivation of $f_{\rm M}\brc{\theta}$ from \eqref{eq:FinalIntegral} tales a long amount of time, and is out of the scope for this manuscript. We hence present the derivations steps and skip the details. Interested readers are referred to \cite[Appendices A-D]{bereyhi2020thesis}.

We start the derivation by taking expectation with respect to noise. This task is done via basic properties of Gaussian integrals. We then use the results in \cite{harish1957differential,itzykson1980planar,guionnet2002large} on the asymptotic limit of \textit{spherical integrals} to calculate the expectation with respect to the sensing matrices. Some short notes on spherical integrals and their asymptotic limits are found in \cite[Apendix E]{bereyhi2020thesis}. Finally, we use the law of large numbers to take the expectation with respect to the true signal samples $\bx^J$.

After taking the expectations, we finally conclude that 
\begin{equation}
	f_{\rm M} \brc{\theta} = \int \exp\set{-N E_{\rm M} \brc{ \mQ^J , \mS^J } +\epsilon_N } \dif \mQ^J \dif \mS^J \label{eq:FinalMoment}
\end{equation}
where the exponent function $E_{\rm M} \brc{ \mQ^J , \mS^J }$ is defined as
\begin{equation}
E_{\rm M} \brc{ \mQ^J , \mS^J } =  \sum_{j=1}^J \dbc{ \mg_j \brc{\mT_j \mQ_j} + \tr{\mS_j \mQ_j} } - \mam \brc{ \mS^J}.
\end{equation}
$\mQ^J$ and $\mS^J$ are further defined as
\begin{subequations}
	\begin{eqnarray}
	\mQ^J &=& \brc{ \mQ_1, \ldots, \mQ_J }\\
	\mS^J &=& \brc{ \mS_1, \ldots, \mS_J }
\end{eqnarray}
\end{subequations}
with $\mQ_j$ and $\mS_j$ being symmetric $\theta \times \theta$ matrices for $j\in \dbc{J}$. The exact definition of integral measures $\dif \mQ_j$ and $\dif \mS_j$ is given in \cite[Appendix A]{bereyhi2020thesis}. $\epsilon_N$ is a bounded sequence in $N$ which converges to zero as $N$ grows large.  $\mT_j$ is defined as
\begin{equation}
	\mT_j = \frac{1}{2\lambda_j} (\mI_\theta - \frac{\beta\sigma_j^2}{\lambda_j + \theta \beta \sigma_j^2} \mone_\theta)
\end{equation}
where $\mI_\theta$ and $\mone_\theta$ denote $\theta\times \theta$ identity and all-one matrices, respectively. The components of the exponent function $E_{\rm M} \brc{ \mQ^J , \mS^J }$ are further defined as follows:
\begin{itemize}
	\item The function $\mg_j \brc{\cdot }$ is given by
		\begin{equation}
			\mg_i\brc{\mM} \coloneqq  \int_{0}^{\beta}  \tr{ \mM \rmR_j \brc{- 2 \mM \omega } } \dif \omega
		\end{equation}
	for a $\theta\times \theta$ matrix $\mM$.
	\item $\mam \brc{ \mS^J}$ is defined as
	\begin{equation}
		\mam \brc{\mS^J} = \Ex{\log \int\limits_{ \bvv_j\in\setX^{\theta} } \exp\set{ \Xi \brc{\bvv^J,\bxx^J \vert \mS^J}   +h \Delta \brc{\bvv^J ; \bxx^J } } \dif \bvv^J }{}
	\end{equation}
where the function $\Xi \brc{\bvv^J,\bxx^J \vert \mS^J}$ is given by
\begin{equation}
\Xi \brc{\bvv^J,\bxx^J \vert \mS^J} = \sum_{j=1}^J \brc{\bxx_j-\bvv_j}^\trp \mS_j \brc{\bxx_j-\bvv_j} -\beta   u_\rmv \brc{\bvv^J}.
\end{equation}
In these equations, $\bvv^J = \brc{\bvv_1,\ldots,\bvv_J}$ and $\bxx^J = \brc{\bxx_1,\ldots,\bxx_J}$, where $\bvv_j \in \setX^\theta$ and $\bxx_j = \xx_j \mone_{\theta\times 1}$ for $j\in\dbc{J}$. $\mone_{\theta\times 1}$ denotes a $\theta \times 1$ all-one vector and $\xx_1,\ldots,\xx_J$ are correlated random variables distributed jointly with $\rmp_X\brc{\xx_1,\ldots,\xx_J}$. It is worth mentioning that the term $u_\rmv \brc{\bvv^J}$ is decomposed as
\begin{equation}
	u_\rmv \brc{\bvv^J} = \sum_{a=1}^\theta u \brc{\rmv_a^J}
\end{equation}
using the decoupling property of the regularization function $u_\rmv\brc{\cdot}$. Here, $\rmv_a^J = \dbc{\rmv_{1a}, \ldots,\rmv_{Ja}}$ with $\rmv_{ja}$ denoting the $a$-th entry of $\bvv_j$.
\end{itemize}
\begin{trailer}{Remark}
	The definition of $f_{\rm M}\brc{\theta}$ contain integrals over $N$-dimensional vectors. These integrals are transformed to integrals over $\theta$-dimensional vectors in the final expression. This transform follows several steps and assumptions. The detailed derivations can be followed in \cite[Appendix A]{bereyhi2020thesis}.
\end{trailer}

\subsection{General Form of the Solution}
The final form of the moment function in \eqref{eq:FinalMoment} enables us to apply the saddle-point method to derive the free energy in the thermodynamic limit. After some lines of derivation, we conclude that the asymptotic distortion is given by
\begin{equation}
	D = \lim_{u\downarrow 0} \lim_{\beta \uparrow \infty} \int\limits_{\bvv_j \in \setX^\theta } \Ex{\Delta \brc{\bvv^J;\bxx^J} \rmq_\beta \brc{\bvv^J \vert \bxx^J} }{} \dif \bvv^J .
\end{equation}
The conditional distribution $\rmq_\beta\brc{\bvv^J  \vert \bxx^J}$ in this equation is a Boltzmann-Gibbs distribution over the reduced dimension and is defined as
\begin{equation}
\rmq_\beta\brc{\bvv^J  \vert \bxx^J} = \dfrac{\exp\set{ -\beta E_0 \brc{\bvv^J,\bxx^J} } }{\displaystyle \int\limits_{\bvv_j\in\setX^\theta} \exp\set{ -\beta E_0 \brc{\bvv^J,\bxx^J} } \dif \bvv^J }
\end{equation}
where the exponent function is defined as
\begin{equation}
	E_0 \brc{\bvv^J,\bxx^J} = \sum\limits_{j=1}^J \brc{\bxx_j - \bvv_j}^\trp \mR_j\brc{\bxx_j  - \bvv_j} + u_\rmv \brc{\bvv^J},
\end{equation}
and the expectation is taken with respect to $\bxx^J$. $\mR_j$ in the exponent function is further defined as
\begin{equation}
	\mR_j \coloneqq \mT_j \rmR_{j} \brc{-2\beta \mT_j \mQ_j^\star}
\end{equation}
where the symmetric $\theta \times \theta$ matrix $\mQ_j^\star$ for $j\in[J]$ is calculated from the following fixed-point equation:
\begin{align}
	\mQ_j^\star &=\int\limits_{\bvv_j\in\setX^\theta }  \Ex{ \brc{\bxx_j  -  \bvv_j} \brc{\bxx_j - \bvv_j}^\trp \rmq_\beta \brc{\bvv^J|\bxx^J} }{ } \dif \bvv^J. \label{eq:FixedPint_General}
\end{align}

\begin{trailer}{Remark}
	To see how \eqref{eq:FixedPint_General} describes a fixed-point equation, note that the conditional distribution $\rmq_\beta\brc{\bvv^J  \vert \bxx^J}$ depends on $\mQ_j^\star$. As the result, the right hand side of this identity is calculated as a function of $\mQ_j^\star$, and \eqref{eq:FixedPint_General} describes a fixed-point equation in $\mQ_j^\star$.
\end{trailer}

\subsection{Constructing Parameterized $\mQ_j^\star$}
The general solution of the replica method is given in terms of $\theta \times \theta$ matrices $\mQ_j^\star$. The reason for having such a solution is simply the \textit{replica continuity} assumption. In this assumption, we postulated that $\theta$ is an \textit{integer}. For an integer $\theta$, having a $\theta \times \theta$ matrix is completely reasonable. Nevertheless, we aim to find the final solution as an analytic function in $\theta$, so that we could use it also for \textit{real} choices of $\theta$.

\begin{svgraybox}
To find an analytic solution, there exists a classic trick: \textit{Assuming an structure on $\mQ_j^\star$}. In this trick, we limit the search to a set of parameterized matrices. The parameterization is considered such that the solution of the fixed-point equation leads to an analytic moment function in $\theta$. 
\end{svgraybox}

In order to clarify this trick, consider the following illustration: We assume that $\mQ_j^\star$ is a $\theta \times \theta$ matrix which is parameterized by $L$ parameters $q^{(1)},\ldots,q^{(L)}$. This means that
\begin{align}
\mQ_j^\star = W_j \brc{q^{(1)}, \ldots, q^{(L)}}
\end{align}
where $W_j \brc{\cdot}$ is a deterministic function which determines a $\theta \times \theta$ matrix for given scalar arguments $q^{(1)},\ldots,q^{(L)}$. Note that $L$ is an integer whose value is fixed and does not vary by changing $\theta$. By inserting this matrix into the fixed-point equation, a system of $L$ coupled equations in terms of $q^{(1)},\ldots,q^{(L)}$ is derived. We insert the solution of this equation system into the replica solution and calculate the limits analytically.

With respect to this trick the following question arises: \textit{What is a meaningful structure for $\mQ_j^\star$?}. The answer to this question is based on physical intuitions and mathematical investigations of an energy model. These discussions are out of the scope of this overview; however, their results can be directly applied to our study.~The investigations in theory of spin glasses suggests a set of recursively extendable structures drawn from the assumption of \textit{\ac{rs}}. These structures start by a simple symmetric parameterization, known as \ac{rs}, and then extends to more advanced structures by iteratively perturbing the \ac{rs}.

\subsubsection{Replica Symmetric Solution}
\ac{rs} considers the most basic structure on $\mQ_j^\star$ which depends only on two parameters $q_j$ and $\chi_j$, and is given by
\begin{align}
	\mQ_j^\star = \frac{\chi_j}{\beta} \mI_\theta + q_j \mone_\theta \label{eq:rs}.
\end{align}
Using this structure, the asymptotic distortion is determined as follows:

\begin{trailer}{Replica Symmetric Solution}
Consider random variables $\xx^J  = \brc{ \xx_1, \ldots, \xx_J}$ distributed with $\rmp_X\brc{\xx^J}$. Let $\yy_j$ for $j\in[J]$ be calculated from $\xx_j$ as 
\begin{equation}
\yy_j  =  \xx_j  +  \zz_j	
\end{equation}
with $\zz_j$ being independent Gaussian random variables with zero mean and variance $\xi_j^2$. The variances $\xi_j^2$ are determined in terms of parameters $\chi_j$ and $q_j$ as
\begin{equation}
	\xi_j^2 = \dbc{ \rmR_j \brc{-\frac{\chi_j}{\lambda_j} } }^{-2} \frac{\partial}{\partial\chi_j} \dbc{ \brc{\sigma_j^2 \chi_j-\lambda_j q_j}\rmR_j \brc{-\frac{\chi_j}{\lambda_j} } }.
\end{equation}
Define $\hxx^J$ as 
\begin{equation}
	\hxx^J = \argmin_{\rmv_j \in \setX } \sum_{j=1}^J \frac{1}{2\tau_j} \brc{ \yy_j-\rmv_j}^2 + u \brc{ \rmv^J}, \label{eq:RS_Estimator}
\end{equation}
where $\tau_j$ is given by 
\begin{equation}
	\tau_j = \frac{\lambda_j}{\rmR_j \brc{-\frac{\chi_j}{\lambda_j} }}.
\end{equation}

The \ac{rs} solution for asymptotic distortion in this case is given by
\begin{equation}
	D = \Ex{\Delta \brc{ \hxx^J ; \xx^J }}{}
\end{equation}
when  $\chi_j$ and $q_j$ are calculated from the following fixed point equations:
\begin{subequations}
	\begin{eqnarray}
		q_j &=& \Ex{ \brc{\hxx_j-\xx_j}^2 }{}\\
		\theta_j^2 \chi_j &=& \tau_j \Ex{ \brc{\hxx_j-\xx_j}  \zz_j}{}.
	\end{eqnarray}
\end{subequations}
The expectation is taken over all random variables, i.e., $\xx^J$ and $\zz^J$.
\end{trailer}

The \ac{rs} solution is calculated readily. In fact, \eqref{eq:RS_Estimator} is a $J$-dimensional optimization which can be solved analytically in various cases. For most well-known \ac{rls}-based recovery algorithms, the \ac{rs} solution gives a valid prediction of the asymptotic distortion. Nevertheless, there are few particular cases in which the \ac{rs} solution is invalid\footnote{In fact, the predicted distortion deviates the known rigorous bounds.}. This inconsistency is due to the simplicity of the \ac{rs} structure. For those cases, one needs to break \ac{rs}.

\subsubsection{Replica Symmetry Breaking}
For scenarios in which the \ac{rs} solution is not valid, the search for $\mQ_j^\star$ is extended to a wider set of parameterized matrices via the \ac{rsb} scheme. This scheme was introduced by Parisi in \cite{parisi1980sequence}. The scheme perturbs the \ac{rs} gradually via a recursive technique. This perturbation is called  \textit{breaking}.

\begin{trailer}{Replica Symmetry Breaking}
	Let $\theta$ be an integer multiple of an integer $\zeta$, and $\mQ_{\ell}$ represent a $\zeta\times\zeta$ matrix. \ac{rsb} finds the new $\theta \times \theta$ matrix $\mQ_{\ell+1}$ as
	\begin{align}
		\mQ_{\ell+1}=  \mI_{\tfrac{\theta}{\zeta}} \otimes \mQ_{\ell} + q_{\ell+1} \mone_\theta
	\end{align}
	for some real scalar $q_{\ell+1}$. Here, $\otimes$ denotes the Kronecker product.
\end{trailer}

By letting $\mQ_0$ to be an \ac{rs} matrix, the \ac{rsb} structures are recursively generated. The \ac{rsb} solutions are of more complicated form. We hence skip them and refer interested readers to \cite[Chapter 4]{bereyhi2020thesis}.

\section{Applications and Numerical Results}
\label{sec:app}

The asymptotic characterization of \ac{rls}-based recovery algorithms enables us to address several tasks which rise in various applications of sparse recovery. In this section, we briefly go through a few of them. The scope of these applications however is not limited to these instances. We have given more discussions in this respect in \cite{bereyhi2020thesis,bereyhi1,bereyhi2,bereyhi3,bereyhi4,bereyhi5,bereyhi6,bereyhi7,bereyhi8,bereyhi9,bereyhi10,bereyhi11,bereyhi12,bereyhi13,bereyhi14,bereyhi15,bereyhi16,bereyhi17}.

\subsection{Performance Analysis of Sparse Recovery}
The most relevant application of the results is to employ them for asymptotic investigation of sparse recovery algorithms. A long discussion in this respect is found in \cite[Chapter 6]{bereyhi2020thesis}, as well as \cite{bereyhi3,bereyhi4,bereyhi7}. As a particular instance, we employ the asymptotic results to study the impact of spatial correlation in multi-terminal compressive sensing.

For sake of visualization, we consider a simple setting with two terminals. These terminals observe signals $x_1\brc{t}$ and $x_2\brc{t}$ which are jointly sparse. We assume that the joint sparsity follows the \textit{common-innovation} model; see Section~\ref{sec:dcs}. 

The fusion center can recover the sparse signal via two approaches:
\begin{enumerate}
	\item Since each signal is sparse, the fusion center can use two separate sparse recovery algorithms to recover each sparse signal \textit{individually}.
	\item A \textit{joint} recovery algorithm can be used to take into account the spatial correlation among the terminals.
\end{enumerate}

From Slepian-Wolf theorem, it is known that the joint recovery can outperform an individual scheme \cite{el2011network}. To investigate this issue, we consider a sample \ac{rls}-based recovery algorithm for each approach and compare their performance using the asymptotic characterization. For the individual approach, we consider the well-known \ac{lasso} algorithm. This algorithm is realized by setting the regularization function to
\begin{equation}
	u_\rmv\brc{\bv_1,\bv_2} = \norm{\bv_1}_1 + \norm{\bv_2}_1.
\end{equation}

As a comparable joint recovery, one can use an \ac{rls}-based joint recovery scheme with convex utility, e.g., the group \ac{lasso} algorithm in which
\begin{equation}
	u_\rmv\brc{\bv_1,\bv_2} = \norm{\bv_1,\bv_2}_{2,1}.
\end{equation}
In the sequel, we use the two-dimensional \ac{lasso} technique proposed initially in \cite{bereyhi7}. This algorithm extends the individual \ac{lasso} recovery approach by modifying the regularization function as
\begin{equation}
	u_\rmv\brc{\bv_1,\bv_2} = \norm{\bv_1}_1 +\norm{\bv_2}_1 + \phi \norm{\bv_1 + \alpha \bv_2}_1
\end{equation}
for some scalars $\phi$ and $\alpha$. The intuition behind this algorithm is that any linear combination of jointly sparse signals is also sparse and its sparsity level depends on the spatial correlation. The study in \cite{bereyhi7} has shown that this approach outperforms the classic group \ac{lasso} technique for the common-innovation joint sparsity model.

\begin{figure}[h]
	\sidecaption[t]
	\input{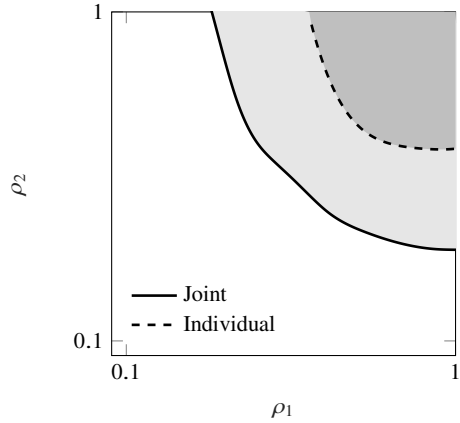}
	\caption{Rate-distortion region for both joint and individual LASSO schemes.}
	\label{fig:1} 
\end{figure}

Using the \ac{rs} solution, we can calculate the asymptotic \ac{mse} for both approaches. The asymptotic \ac{mse} is determined from the \ac{rs} solution by setting the distortion function to the squared Euclidean distance between the true and recovered pairs. Using the asymptotic \ac{mse}, we plot the \textit{rate-distortion} region for both schemes. It is found by fixing a threshold \ac{mse} and finding all pairs of compression rates, i.e., $\brc{\rho_1,\rho_2}$ for which the achievable \ac{mse} is smaller that the threshold. This region is shown in Fig.~\ref{fig:1} for a particular example in which the common part is $30\%$ sparse and each terminal has a $10\%$ sparse innovation component. The tunable factors in both algorithms are optimized to achieve minimal \ac{mse}. As the figure shows, using a spatially coupled regularization improves the recovery performance significantly. The Bayesian viewpoint illustrates this observation as follows: The postulated prior distribution of an \ac{rls}-based algorithm with spatially coupled regularization takes into account the spatial correlation, and hence outperforms the individual approach.

The asymptotic characterization can be further used to investigate those recovery algorithms whose performances are unknown. An example is a non-convex $\ell_{p,q}$-norm minimization, e.g.,
\begin{equation}
	u_\rmv\brc{\bv_1,\bv_2} = \norm{\bv_1,\bv_2}_{2,q}
\end{equation}
with $0 \leq q\leq 1$. These algorithms are not computationally tractable, and hence their performance is left unknown. The asymptotic characterization via the replica method can give a prediction on the performance of these algorithm. More discussions can be followed in \cite[Chapter 6]{bereyhi2020thesis}.

\subsection{Tuning RLS-Based Algorithms}
Compressive sensing is not the only application of sparse recovery. In fact, sparse recovery is used in various applications, such as communications, networking and machine learning; see some instances in \cite{bereyhi2020thesis,bereyhi2,bereyhi5,bereyhi6,bereyhi8,bereyhi9,bereyhi10,bereyhi11,bereyhi12,bereyhi13,bereyhi14,bereyhi15,bereyhi16,bereyhi17,bereyhi2Extension}. In these applications, there is often a tuning task: Find the \textit{regularizers} of an \ac{rls}-based algorithm, such that the performance is optimized. This task is readily addressed via the asymptotic characterization of the \ac{rls}-based recovery algorithms.

We can illustrate this application by considering a simple example of \textit{spatial modulation}. The details on this example can be followed in \cite{bereyhi13,bereyhi17}. In spatial modulation, the information is encoded in the support of the transmit signal: In each symbol interval, based on the data bits, a subset of available transmit antennas is set on and the remaining are turned off. As a result, the transmit signal is sparse, and hence an effective detection scheme at the receiver is to use a sparse recovery algorithms\footnote{For sake of brevity, we skip the detailed system model. Interested readers are referred to \cite{bereyhi13,bereyhi17} and the references therein.}.

The common sparse recovery algorithms used in \textit{spatial modulation} are formulated as \ac{rls}-based recovery schemes. Examples are the classic \ac{lasso} and \textit{box-\ac{lasso}} techniques. We already know the classic \ac{lasso} scheme from the previous section. The \textit{box-\ac{lasso}} technique is moreover an extension of \ac{lasso} in which the set $\setX$ is restricted to a box, e.g., $\setX = \dbc{-B,B}$ for some real $B$. This box restriction is shown to enhance the performance, when we detect discrete-valued signals.

\begin{figure}[h]
	\sidecaption[t]
	\input{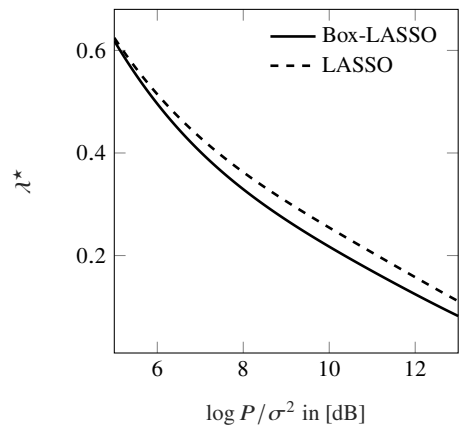}
	\caption{Optimal regularizer for LASSO and box-LASSO.}
	\label{fig:2} 
\end{figure}

One of the challenges in these techniques is to find the optimal regularizers, which results in minimum bit error rate. Such a task is usually addressed via iterative tuning techniques. Nevertheless, in high data rates, the tuning techniques impose extra processing load on the system. The asymptotic characterization enables us to address this task analytically, and hence avoid the extra load. An instance of tuning via the asymptotic characterization is shown in Fig.~\ref{fig:2}. In this figure, a multiuser uplink scenario is considered in which the \ac{lasso} and box-\ac{lasso} techniques are used for detection. Here, $P$ denotes the transmit power and $\sigma^2$ is the noise variance~at the receiver. The sparsity of the transmit signal is assumed to be $12.5 \%$. The figure shows the optimal regularizer, denoted by $\lambda^\star$ against $\log P/\sigma^2$. Although these results are derived via the asymptotic characterization, the study in \cite{bereyhi17} shows that they closely track the simulation results. Further discussions regarding the tuning of \ac{rls}-based algorithms via asymptotic results can be followed in \cite[Chapters 6 and 7]{bereyhi2020thesis}, as well as \cite{bereyhi10}.

\section{Summary and Final Discussions}
The replica method is a powerful tool for large system analyses, as seen this in this manuscript. Following the prescription suggested by the replica method, we have found an analytic expression for the asymptotic distortion. The result could not be derived via basic analytical tools. This demonstrates the power of the \textit{replica method}. To keep the contents of this manuscript straightforward, we have dropped the detailed derivations and only presented the major steps. The details can be followed in \cite{bereyhi2020thesis}.

The presented analysis is extendable in various respects, and results in various further interesting conclusions. Going through all of these extensions and conclusions is not possible within a short manuscript. We hence skip them here and refer interested readers to \cite{bereyhi2020thesis} and the references therein. Nevertheless, to give you a flavor, we conclude this manuscript by giving a few highlights.

\subsection{Decoupling Principle}
Although this manuscript focused on the derivation of \textit{asymptotic distortion}, the result can be further used to prove the so-called \textit{decoupling principle}. This principle indicates that in the asymptotic regime the joint distribution of $x_n^J$ and $\hx_n^J$ converges to the one described via an \textit{equivalent} scalar system, often called the \textit{decoupled system}. 
This decoupled setting is shown to consist of an equivalent additive noise term and a decoupled recovery scheme; see \cite[Chapter 5]{bereyhi2020thesis}. The interesting point is that the decoupled recovery scheme remains the same for all solutions, i.e., the RS and RSB solutions, and it is only the distribution of the equivalent noise term which changes. A comprehensive illustration of the decoupling principle and its detailed derivations are given in \cite[Chapter 5]{bereyhi2020thesis}.

\subsection{Nonuniform Sparsity Patterns}
In various applications, the sparsity of signals varies over time. This form of sparsity is often called \textit{nonuniform}, whereas the normal form is considered \textit{uniform}. For nonuniform sparse signals, the stochastic model of samples is not \ac{iid} anymore. They are still independent\footnote{Since temporal correlation is usually avoided by classic sampling approaches.}; however, the joint distribution changes through time. The analysis in this manuscript extends to nonuniform patterns by some modifications. Some results on this direction can be followed in \cite{bereyhi1Extension,bereyhi14}.

\subsection{Extensions to Bayesian Estimation}
In the Bayesian framework, the considered \ac{rls}-based algorithms are seen as \ac{map} estimators. This is however not the only approach for Bayesian inference. In many other applications, e.g., signal processing and machine learning, other forms of Bayesian inference are used, e.g., the minimum \ac{mse} estimator or more generally estimators with minimal posterior distortion; see for example \cite{bereyhi2Extension}.

The replica based analysis in this manuscript is readily extended to these estimators as well. The derivations follow the same steps as illustrated in this manuscript, i.e., finding a corresponding spin glass and interpreting the desired metrics as its macroscopic parameters. The key difference here is that for other estimators, the desired metrics might be a macroscopic parameter at a \textit{non-zero} temperature.

\section{Bibliographical Notes}
Primary studies on asymptotic analysis were limited to linear recovery techniques, e.g., studies in \cite{guo1999linear,shamai2001impact}. M\"uller and Gerstacker conjectured later that similar behavior extends to most nonlinear schemes, as well \cite{muller2004capacity}. This conjecture was originated from the analytic results reported in a series of studies which employed the replica method to derive the asymptotic performance of multiuser detectors. This line of work started with the study by Tanaka in \cite{tanaka2002statistical}. A key milestone in this direction was achieved in \cite{guo2005randomly}, where the authors determined the asymptotic performance metrics of a mismatched minimum \ac{mse} recovery scheme. This result was later extended to \ac{map} estimators in \cite{rangan2012asymptotic} using standard large deviations techniques. 

With respect to the problem of sparse recovery, most initial asymptotic analyses relied on the replica analyses presented in \cite{tanaka2002statistical} and \cite{guo2005randomly}. This led to results which enclosed restricted system models, e.g., single terminal and \ac{iid} sensing matrix. The later lines of work deviated from this approach and used the replica method explicitly to derive the asymptotic characteristics; for instance \cite{tulino2013support,vehkapera2014analysis,kabashima2010statistical,kabashima2009typical,wen2016sparse}. These analyses were however limited to \ac{rs} investigations. The complete replica analysis of \ac{rls}-based algorithms was given in a series of work in \cite{bereyhi4,bereyhi3,bereyhi1,bereyhi2020thesis} providing both \ac{rs} and \ac{rsb} solutions.

\bibliographystyle{spmpsci}
\bibliography{bibfileLG.bib}

\end{document}